\documentclass[aps,pra,twocolumn,showpacs,floatfix,superscriptaddress]{revtex4-1}
\usepackage{amsmath,amssymb,amsfonts,bbm,graphicx,hyperref,times}
\usepackage{color}      

\def\Tr{\hbox{Tr}}
\def\sigmaCM{\boldsymbol{\sigma}}
\newcommand{\be}{\begin{equation}}
\newcommand{\ee}{\end{equation}}
\newcommand{\bea}{\begin{align}}
\newcommand{\eaa}{\end{align}}

\newcommand{\sig}{{\boldsymbol{\sigma}}}

\begin{document}
\title{Cram\'er-Rao bound for time-continuous measurements in linear Gaussian quantum systems}
\author{Marco G. Genoni}
\affiliation{Quantum Technology Lab, Dipartimento di Fisica, Universit\`a degli Studi di Milano, 20133 Milano, Italy}
\begin{abstract}
We describe a compact and reliable method to calculate the Fisher information for the estimation of a dynamical parameter in a continuously measured linear Gaussian quantum system. Unlike previous methods in the literature, which involve the numerical integration of a stochastic master equation for the corresponding density operator in a Hilbert space of infinite dimension, the formulas here derived depends only on the evolution of first and second moments of the quantum states, and thus can be easily evaluated without the need of any approximation. We also present some basic but physically meaningful examples where this result is exploited, calculating analytical and numerical bounds on the estimation of the squeezing parameter for a quantum parametric amplifier, and of a constant force acting on a mechanical oscillator in a standard optomechanical scenario. 
\end{abstract}
\maketitle
\section{Introduction}
Parameter estimation via quantum probes and quantum measurements will 
lead to a new generation of detectors characterised by sensitivities not achievable through
only classical means \cite{GiovannettiNatPhot,RafalPO}. 
The promised quantum enhancement is typically however lost as soon as some decoherence affects the system \cite{EscherNatPhys,KolodynskyNatComm}. On the other hand, the information leaking 
into the environment can be in principle used for parameter estimation as well, in particular via time-continuous monitoring of the environment itself \cite{WisemanMilburn,SteckJacobs}. While several strategies based on time-continuous measurements and feedback have been proposed for quantum state engineering, in particular with the main goal of generating steady-state squeezing and entanglement \cite{WisemanMilburn,WisemanSqueezing,WisemanMancini,SerafozziMancini,DohertyPRL2011,Tempura,Ravotto,Hofer2015,Levante,Calypso} or to 
study and exploit trajectories of superconducting qubits \cite{MolmerMurch2015,MolmerMurch2016}, less attention has been devoted to parameter estimation. Notable exceptions are the estimation of a magnetic field via a continuously monitored atomic ensemble \cite{GeremiaMagnetometry2003}, the tracking of a varying phase
\cite{BerryWiseman2008,MankeiWaveform2011,SciencePhTracking2012}, the estimation of Hamiltonian and environmental parameters \cite{Ang2013, Iwasawa2013,GammelmarkCRB,GammelmarkQCRB,KiileyrichHomodyne,MankeiSpectrum2016,Guta2016,Buchman2016} and optimal state estimation for a cavity optomechanical system \cite{KalmanOpto}.\\
The ultimate precision achievable by quantum metrological strategies is determined by the classical and  quantum Cram\`er-Rao bounds \cite{HelstromBook,CavesBraunstein,MatteoIJQI}, which are expressed in terms of, respectively, 
classical and quantum Fisher information (FI). Recently, methods have been proposed to calculate 
these quantities in the stationary regime for certain relevant setups \cite{Ang2013,Iwasawa2013,Guta2016}, or in the dynamical regime in the case of time-continuous homodyne and photon-counting measurements \cite{GammelmarkCRB,GammelmarkQCRB}. In particular, in order to evaluate
the FI corresponding to a continuous homodyne detection, the method presented in Ref.\cite{GammelmarkCRB} relies on the integration of stochastic master equations for operators characterizing the quantum state and the measurement performed. While this can be straightforwardly accomplished in the case of finite-dimensional quantum systems, such as two-level atoms and superconducting qubits, the method becomes computationally very expensive and less reliable in the case of large or even infinite-dimensional systems, such as the electromagnetic field, atomic ensembles and mechanical oscillators. In fact, in these cases, one has to truncate the corresponding Fock space, posing a constraint on the maximum energy of the system. \\
The goal of this article is to provide an efficient and reliable method to calculate the FI for parameter estimation via time-continuous measurements in linear Gaussian quantum systems. Gaussian systems represents a subclass of infinite-dimensional bosonic systems, whose properties and dynamics can be univocally described in terms of first and second moments only \cite{GaussianAOP,GaussianRMP,diffusone}. In order to observe such a dynamics, one has to consider Hamiltonians at most quadratic in the canonical operators, a linear coupling with the environment and time-continuous monitoring via Gaussian measurements \cite{diffusone,WisemanDiosi,WisemanDoherty}. This restriction allows one to greatly simplify the analysis of infinite-dimensional quantum systems, and at the same time describe several state-of-the-art experimental setups in the area of quantum optics, opto-mechanics, trapped ions and atomic ensembles. In particular, we remind the reader how optimal state estimation via time-continuous monitoring has been very recently accomplished for a Gaussian cavity optomechanical systems \cite{KalmanOpto}, showing the timeliness and relevance of this approach. \\
In detail, the manuscript is structured as follows. In Sec. \ref{s:gauss} we provide a basic introduction on Gaussian systems and their diffusive and conditional dynamics, while in Sec. \ref{s:fisher} we revise the Cram\'er-Rao bound, with a focus on {\em a posteriori} Gaussian distributions. In Sec. \ref{s:cramerrao} we present the main result of the manuscript, that is a method for the calculation of the FI for Gaussian systems depending only on the evolution of first and second moments and that does not need any approximation or limit on the energy of the quantum states in exam. To show the potential of our results, in Sec. \ref{s:examples} we provide two examples, calculating numerical and analytical bounds on the estimation precision for the squeezing parameter in a quantum parametric amplifier, and for a constant force acting on a mechanical oscillator in a standard opto-mechanical setup. 
\section{Diffusive and conditional dynamics in linear Gaussian systems} \label{s:gauss}
We consider a set of $n$ bosonic modes described by a vector of quadrature operators $\hat{\bf r}^{\sf T}=(\hat{x}_1,\hat{p}_1,\dots,\hat{x}_n,\hat{p}_n)$, satisfying the canonical commutation relation $[\hat{\bf r}, \hat{\bf r}^{\sf T}] = i \Omega$  (with $\Omega_{jk}=\delta_{k,j+1} - \delta_{k,j-1}$) \cite{noteOuterProduct}. We define a quantum state $\varrho$ Gaussian, if and only if can be written as a ground or thermal state of a quadratic Hamiltonian, {\em i.e.} 
\begin{align}
\varrho = \frac{\exp\{-\beta \hat{\mathcal{H}}_{\sf G}\}}{Z} \:, \:\:\: \beta \in \mathbbm{R}\:,
\end{align}
where $\hat{\mathcal{H}}_{\sf G}= (1/2) \hat{\bf r}^{\sf T} H_G \,\hat{\bf r}$ and $H_G \geq 0$ \cite{GaussianAOP,diffusone}. Gaussian states can be univocally described by the vector of first moments ${\bf R}$ and the covariance matrix ${\boldsymbol \sigma}$ \cite{noteOuterProduct}:
\begin{align}
{\bf R} = \Tr[ \varrho \hat{\bf r}]\:,  \qquad
\sigmaCM = \Tr[\varrho \{ \hat{\bf r} - {\bf R} , (\hat{\bf r} - {\bf R})^{\sf T} \} ] \:.
\end{align}
We also recall that, in order to describe a proper Gaussian quantum state, the covariance matrix has to satisfy the physicality condition $\sigmaCM + i \Omega \geq 0$ \cite{GaussianSimon}. 

We now consider a dynamics generated by a Hamiltonian with linear and quadratic terms of the form
\begin{align}
\hat{\mathcal{H}}_s = \frac{1}{2} \hat{\bf r}^{\sf T} H_s \hat{\bf r} - \hat{\bf r}^{\sf T}\Omega {\bf u} \:, \label{eq:Hs}
\end{align}
where $H_s$ is a matrix of dimension $2n \times 2n$, while ${\bf u}$ is a $2n$-dimensional vector.
We also assume that the system is coupled to a large Markovian environment described by a {\em train} of incoming modes $\hat{\bf r}_{b}(t)$, each of which interacts with the system at a given time $t$. The correlations characterizing the environment are specified through the white-noise condition 
\begin{align}
\langle \{ \hat{\bf r}_b (t), \hat{\bf r}_b^{\sf T} (t') \} \rangle = \sigmaCM_b \delta(t-t') \:, \:\:\:\: \sigmaCM_b + i \Omega \geq 0.
\end{align}
The interaction in a interval of time $dt$ between the system and the environment is ruled by the Hamiltonian $$ \mathcal{\hat{H}}_C \,dt = \hat{\bf r}^{\sf T} C\, \hat{\bf r}_b(t) \, dt = \hat{\bf r}^{\sf T} C\, {\rm d}\hat{\bf r}_b(t) = \hat{\bf r}^{\sf T} C \,\hat{\bf r}'_b(t) \, dW.  $$
Here we have introduced the so-called quantum Wiener increment \cite{gardiner} 
$$ {\rm d}\hat{\bf r}_b(t)= \hat{\bf r}_b(t) \, dt = \hat{\bf r}'_b(t) \,dW, $$  
with $dW$ being a real Wiener increment such that $dW^2=dt$, and where
$\hat{\bf r}_b'(t)$ is a vector of ``proper''
dimensionless field operators (that can be associated with {\em detector clicks} in the laboratory, and formally with POVM operators in the Hilbert space) satisfying the canonical commutation relation $[\hat{\bf r}_b'(t),\hat{\bf r}_b'(t)^{\sf T}]=i \Omega$ (we refer to Ref. \cite{diffusone} and to Appendix \ref{appendix2} for more details on these definitions and on the derivation of the following formulas). By tracing out the degrees of freedom of the environment, the dynamics of the Gaussian state is then described by the following equations 
\begin{align}
\frac{d{\bf R}_t}{dt} &= A {\bf R}_t + {\bf u} \\
\frac{d\sigmaCM_t}{dt} &= A \sigmaCM_t + \sigmaCM_t A^{\sf T} + D
\end{align}
where we have introduced the drift matrix $A = \Omega H_s + (\Omega C \Omega C^{\sf T})/2$ and the diffusion matrix $D=\Omega C \sigmaCM_b C^{\sf T} \Omega^{\sf T}$. One should notice that, as expected, the linear term in the Hamiltonian (\ref{eq:Hs}) is responsible for only a displacement of the first moments vector, while the evolution of the covariance matrix is not affected. 
\\
We now assume that the environment is continuously monitored at each time via a Gaussian measurement described by a matrix $\sigmaCM_m$ (s.t. $\sigmaCM_m + i \Omega \geq 0$ ), and whose measurement outcome corresponds to a vector ${\bf x}_m$. In this scenario, the conditional state is still Gaussian, and the dynamics is described by a stochastic equation for the first moment vector and by a deterministic Riccati equation for the covariance matrix \cite{diffusone} (see also Appendix \ref{appendix2}):
\begin{align}
d{\bf R}_t &= A {\bf R}_t \, dt + {\bf u} \, dt + \left(\frac{\sigmaCM_t B + N}{\sqrt{2}}\right) {\bf dw}\:, \nonumber \\
\frac{d\sigmaCM_t}{dt} &= A \sigmaCM_t + \sigmaCM_t A^{\sf T} + D -  (\sigmaCM_t B + N ) (\sigmaCM_t B + N )^{\sf T} \:, \label{eq:Kalman}
\end{align}
where ${\bf dw}$ is a vector of independent Wiener increments (s.t. $dw_j dw_k = \delta_{jk} dt$) and we have introduced the matrices $B = C \Omega (\boldsymbol{\sigma}_b + \boldsymbol{\sigma}_m )^{-1/2}$ and $N = \Omega C \sigmaCM_b (\sigmaCM_b + \sigmaCM_m)^{-1/2}$. \\
The dynamics we have just presented in terms of first and second moments for Gaussian states, can be equivalently described by the following family of stochastic master equations for the (infinite-dimensional) density operator
\begin{align}
d\varrho = - i [ \mathcal{\hat{H}},\varrho]\, dt + \sum_{j=1}^L \mathcal{D}[\hat{c}_j] \varrho \, dt + d{\bf z}^\dagger \Delta \hat{\bf c} \varrho + \varrho \Delta \hat{\bf c}^\dagger d{\bf z} \:,
\end{align}
where $\hat{\bf c} = \widetilde{C} \hat{\bf r}$, $\mathcal{D}[\hat{o}]\varrho = \hat{o}\varrho \hat{o}^\dag - \{ \hat{o}^\dag \hat{o} , \varrho \}/2$, $\Delta \hat{o} = \hat{o} - \Tr[\varrho \hat{o}]$ and $d{\bf z}$  is a vector of complex Wiener increments \cite{WisemanDiosi,WisemanDoherty}.
\\
For our purposes is important to recall that the outcomes ${\bf x}_m$ of the measurement performed on the bath operators $\hat{\bf r}_b'(t)$ are distributed according to a Gaussian multi-variate distribution with mean value $\bar{\bf x}_m = \Omega C^{\sf T} {\bf R}_t \, dW$ 
and covariance matrix $(\sigmaCM_b + \sigmaCM_m )/2$. Typically, the results of time-continuous measurements are formulated as a real current with uncorrelated noise \cite{WisemanDoherty}, {\em i.e.}
\begin{align}
d{\bf y} &:= (\sigmaCM_b + \sigmaCM_m)^{-1/2} \:{\bf x}_m \,dW \\
 &= - B^{\sf T} {\bf R}_t \: dt +\frac{\bf dw}{\sqrt{2}} \:. \label{eq:current}
\end{align}
%

\section{Fisher information from a multivariate Gaussian distribution} \label{s:fisher}
Let us  consider an {\em a posteriori} distribution $p({\bf x} | \theta )$, where the vector ${\bf x}$ corresponds to the outcomes of the measurement performed and $\theta$ is a parameter we want to estimate. In particular we assume $p({\bf x} | \theta)$ to be a multi-variate Gaussian distribution with mean value $\bar{\bf x}_\theta$ and covariance matrix $\Sigma$, where only the mean value depends on the parameter $\theta$. The ultimate limit on how accurate we can estimate $\theta$ is determined by the Cram\`er-Rao bound \cite{CRBound}
\begin{align}
{\rm Var}(\hat{\theta}) \geq \frac{1}{M F(\theta)} \:, \label{eq:CRB}
\end{align}
where ${\rm Var}(\hat{\theta})$ is the variance of an unbiased estimator $\hat{\theta}$, $M$ is the number of measurements, and 
\begin{align}
F(\theta) = \mathbbm{E}_p \left[ \left( \frac{\partial \ln p({\bf x} | \theta )}{\partial \theta} \right)^2 \right]
\end{align}
is the FI corresponding to the distribution $p({\bf x} |\theta)$. As it is clear from the inequality (\ref{eq:CRB}), the FI $F(\theta)$ quantifies how well we can infer the value of the parameter $\theta$ from the measurement outcomes.
By explicitly writing the square of the derivative of the likelihood function $l({\bf x}|\theta)=\log p({\bf x}|\theta)$ and exploiting the property 
$
\mathbbm{E}_p[({\bf x} - \bar{\bf x}_\theta)_j ({\bf x} - \bar{\bf x}_\theta)_k ] = \boldsymbol{\Sigma}_{jk} \:,
$
one can easily prove that the FI corresponding to a Gaussian {\em a posteriori} distribution reads
\begin{align}
F(\theta) = (\partial_\theta \bar{\bf x}_\theta )^{\sf T} \boldsymbol{\Sigma}^{-1} (\partial_\theta \bar{\bf x}_\theta )
\label{eq:FisherGauss}
\end{align}
\section{Fisher Information for time-continuous measurements in linear Gaussian quantum systems}\label{s:cramerrao}
Let us assume that we want to estimate the value of a parameter $\theta$ that characterizes the dynamic of a Gaussian linear quantum system described by Eqs. (\ref{eq:Kalman}). In particular we also assume that only the system Hamiltonian $\hat{\mathcal{H}}_s$ depends on $\theta$, and, thus only the drift matrix $A_\theta$ and/or the vector ${\bf u}_\theta$ depends on the parameter \cite{noteExtension}.\\
We stated above that the probability distribution $p({\bf x}_m | \theta)$ corresponding to the measurement performed at time $t+dt$ on the environment is a Gaussian distribution centered in $\Omega C^{\sf T} {\bf R}_t \, dW$, and with covariance matrix $(\sigmaCM_b + \sigmaCM_m)/2$. As only the mean value depends on the parameter $\theta$, via the first moment vector ${\bf R}_t$, we can exploit Eq. (\ref{eq:FisherGauss}) and write the corresponding infinitesimal FI as
\begin{align}
{dF}_t^{\sf (traj)} (\theta) = 2 (\partial_\theta {\bf R}_t)^{\sf T} C \Omega^{\sf T} (\sigmaCM_b + \sigmaCM_m)^{-1} \: \Omega C^{\sf T} (\partial_\theta {\bf R}_t) \: dt \:. \label{eq:FisherTCM}
\end{align}
Notice that this FI corresponds to a specific trajectory, as it is calculated via the vector $(\partial_\theta {\bf R}_t)$ whose evolution is in principle stochastic and described by the equations
\begin{widetext}
\begin{align}
d(\partial_\theta {\bf R}_t) &= (\partial_\theta A) {\bf R}_t \: dt + A (\partial_\theta {\bf R}_t) \: dt + (\partial_\theta {\bf u})\: dt + \frac{(\partial_\theta \sigmaCM_t) B}{\sqrt{2}} \: {\bf dw}
+  \left(\frac{\sigmaCM_t B + N}{\sqrt{2}}\right) (\partial_\theta {\bf dw})
 \:, \nonumber \\
\frac{d(\partial_\theta \sigmaCM_t)}{dt} &= (\partial_\theta A)\sigmaCM_t + \sigmaCM_t (\partial_\theta A)^{\sf T} + 
A (\partial_\theta \sigmaCM_t) + (\partial_\theta \sigmaCM_t) A^{\sf T} - (\partial_\theta \sigmaCM_t) B (\sigmaCM_t B + N)^{\sf T} - (\sigmaCM_t B + N )((\partial_\theta \sigmaCM_t) B)^{\sf T}, \label{eq:derivatives}
\end{align}
\end{widetext}
where, from Eq. (\ref{eq:current}), we obtain that $\partial_\theta {\bf dw}= \sqrt{2} B^{\sf T} (\partial_\theta {\bf R}_t ) \, dt $.
As a consequence, the actual FI for the measurement performed at time $t+dt$ is evaluated by averaging over all the possible trajectories, {\em i.e.}:
\begin{align}
dF_t (\theta) = \mathbbm{E}_{\bf dw} [ dF_t^{\sf (traj)} (\theta) ] \:. \label{eq:average}
\end{align}
Finally, if one consider the whole data set, {\em i.e.} the continuous stream of measurement outcomes ${\bf d} = \{ {\bf x}_m \}_{t'=0}^{t}$ obtained up to time $t$, the FI corresponding to the whole {\em a posteriori} distribution $p({\bf d}|\theta)$ can be calculated, exploiting its additive property \cite{footnoteADD}, by integrating it (numerically or analytically) as
\begin{align}
F_t(\theta) =\int_{t'=0}^t dF_{t'}(\theta). \: \label{eq:integral}
\end{align}
\\

\section{Examples}\label{s:examples}
\subsection{Estimation of the squeezing parameter for a quantum parametric amplifier} 
\begin{figure}[t!]
\begin{center}
\includegraphics[width=0.95\columnwidth]{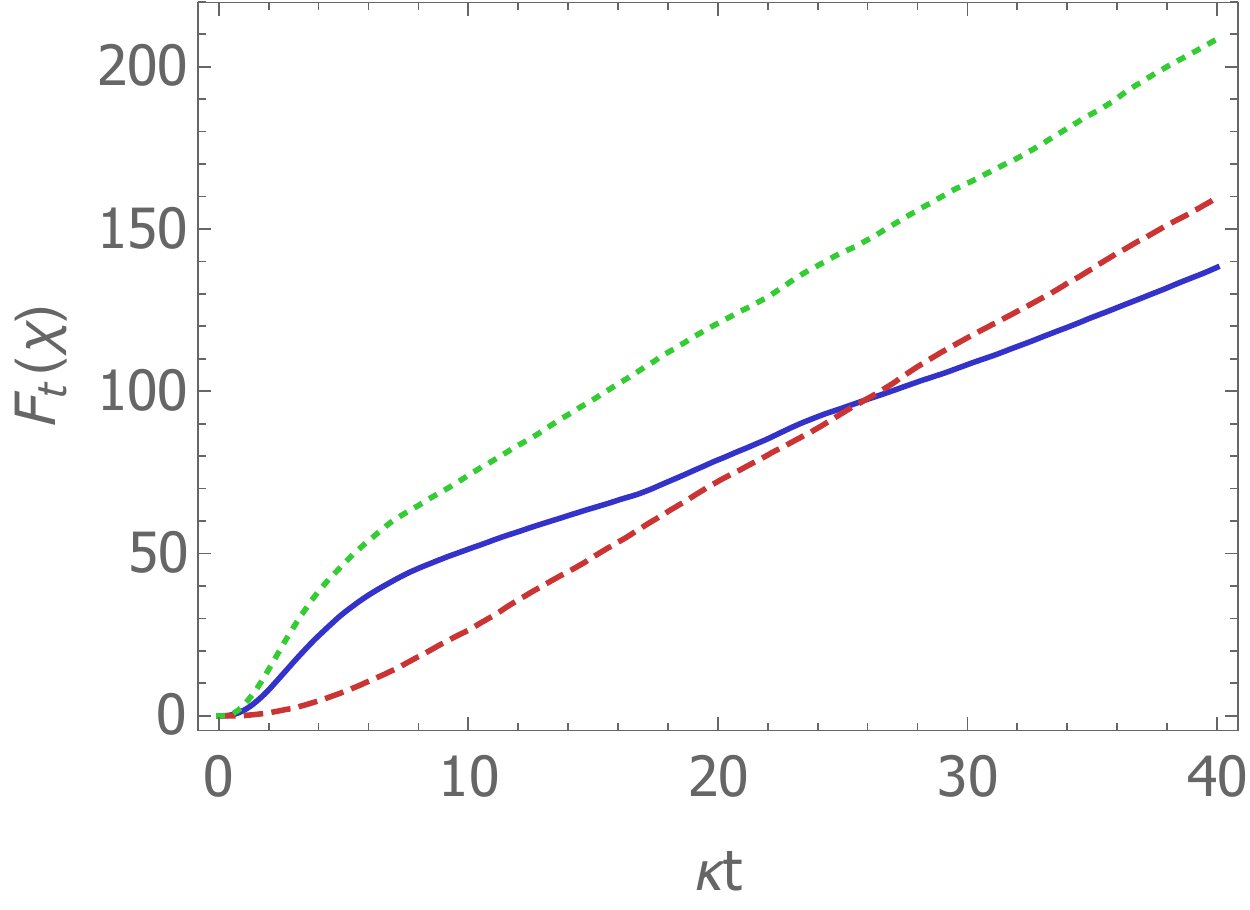} \hspace*{.2cm}
\end{center}
\caption{
FI $F_t(\chi)$ for continuous homodyne and heterodyne detection as a function of time and for $\chi = -0.2 \kappa$ (numerical evaluation with $2000$ trajectories). \\
Green dotted line: homodyne detection of quadrature $\hat{x}$; red dashed line: homodyne detection of quadrature $\hat{p}$; blue solid line: heterodyne detection.
\label{f:squeezest}}
\end{figure}

Let us consider the Hamiltonian for a degenerate parametric amplifier, which in the cavity mode rotating frame reads $\hat{\mathcal{H}}_s = - \chi (\hat{x}\hat{p} + \hat{p}\hat{x})/2$, and assume that the corresponding cavity mode is weakly interacting with a bath at zero temperature, such that the interaction matrix corresponds to a beam-splitter with $C=\sqrt{\kappa} \Omega$ and the correlations of the bath are described by $\sigmaCM_b = \mathbbm{1}_2$. The corresponding master equation for the density operator reads
$\dot{\varrho} 
= - i [ \hat{\mathcal{H}}_s,\varrho] + \kappa\mathcal{D}[\hat{a}] \varrho 
$
where $\hat{a}= (\hat{x} + i \hat{p})/\sqrt{2}$ is the bosonic annihilation operator. In our formalism the dynamics is described by the following drift and diffusion matrices: $A = {\rm diag}(-\chi - \kappa/2 , \chi - \kappa/2 )$, $D=\kappa \mathbbm{1}_2$. 
\\
We now assume to perform a time-continuous measurement on the cavity output, i.e. on the environmental modes after the interaction with the cavity mode, via a Gaussian measurement, in order to estimate the squeezing coupling constant $\chi$. In our analysis we will focus on two type of measurements: a time-continuous homodyne measurement of a quadrature $\hat{r}(\phi) = (\cos\phi\: \hat{x} + \sin\phi \: \hat{p})/\sqrt{2}$ and a time-continuous heterodyne detection, i.e. the projection on single-mode coherent states (the details on the Gaussian description of these measurements can be found in Appendix \ref{appendix1}).Under these assumptions, Eq. (\ref{eq:FisherTCM}) for the infinitesimal FI simplifies to the following equations
\begin{align}
dF_{t} &= 2 \kappa \: \mathbbm{E}_{\bf dw} [ (\partial_\chi \langle \hat{r}_\phi \rangle_t )^2 ] \,dt \:\:\: \textrm{homodyne quadrature} \: \hat{r}_\phi  \nonumber \\
dF_{t} &= \kappa \:  \mathbbm{E}_{\bf dw} [(\partial_\chi \langle \hat{x} \rangle_t )^2 +  (\partial_\chi \langle \hat{p} \rangle_t )^2 ] \,dt \:\:\: \textrm{heterodyne}  \nonumber
\end{align}
In Fig. \ref{f:squeezest} we plot the FI $F_t(\chi)$ as a function of time. We notice that the best performances, between the three measurements here considered, are obtained via a time-continuous homodyne measurement of the quadrature $\hat{x}$, which, for the parameters we have chosen, is the quadrature being anti-squeezed by the Hamiltonian 
$\hat{\mathcal{H}}_s$. One can also observe how at long times, all three curves present a linear behaviour, indicating that the infinitesimal FI takes the form $dF_t (\chi)= K \,dt$, and that the homodyne monitoring of $\hat{p}$ overcomes the performance of heterodyne detection.
\subsection{Estimation of a constant force on a mechanical oscillator}
\begin{figure}[t!]
\begin{center}
\includegraphics[width=0.95\columnwidth]{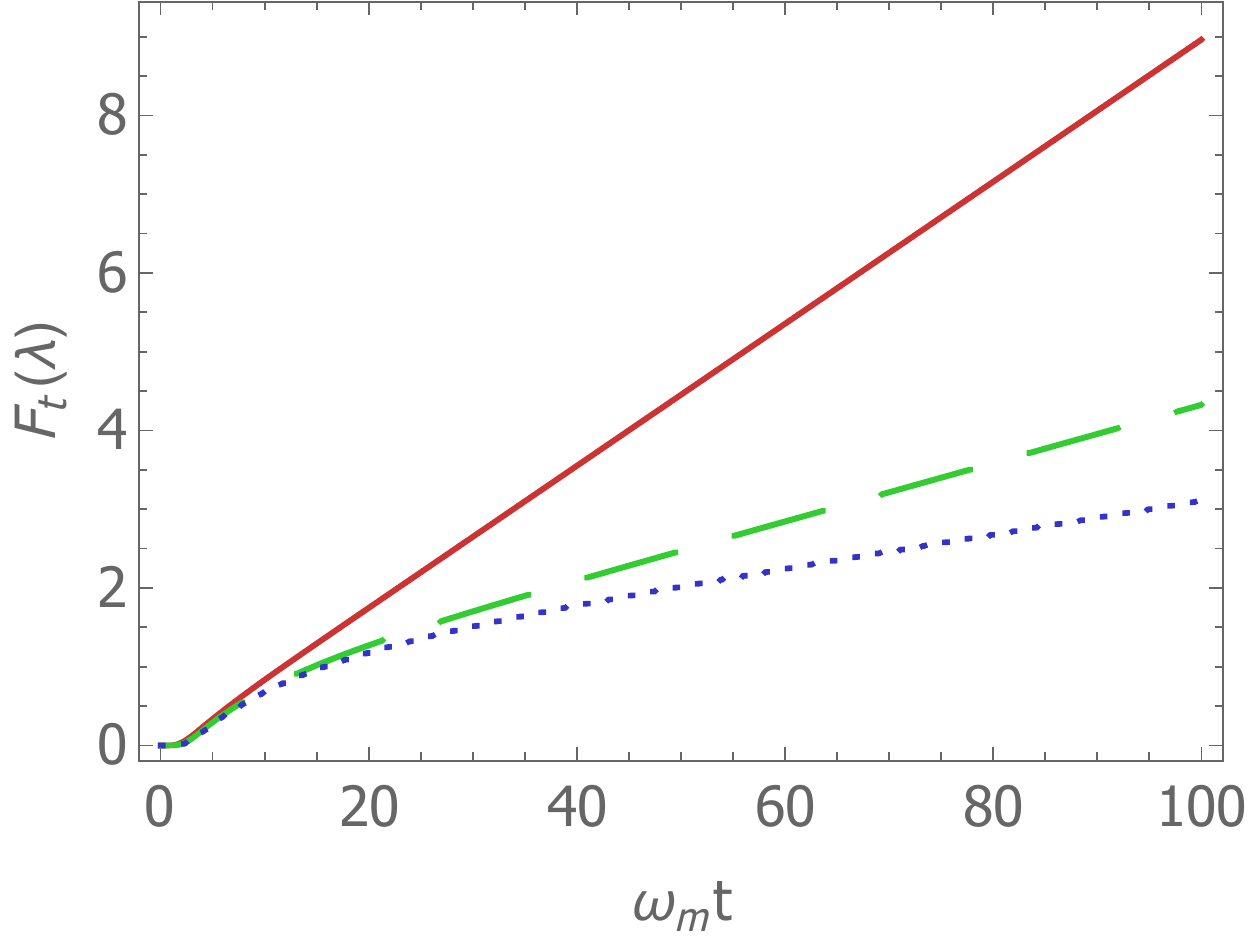} \hspace*{.2cm}
\end{center}
\caption{
FI $F_t(\lambda)$ for continuous homodyne detection of the cavity field quadrature $\hat{p}_c$ as a function of time and for different values of the cavity decay rate: $\kappa=\omega_m/2$ - red solid line; $\kappa = \omega_m/10$ - green dashed line; $\kappa = \omega_m/20$ - blue dotted line (the other parameters are chosen as: $g=\omega_m/2$, $\gamma=\omega_m/3$, $\eta=1$). The inset shows the behaviour of the FI at small times.
\label{f:forceest}}
\end{figure}

 We consider a standard cavity optomechanical setup where a mechanical oscillator oscillating at frequency $\omega_m$, is coupled to a cavity mode characterized by a resonance frequency $\omega_c$ and driven with a laser at frequency $\omega_l = \Delta + \omega_c$. The interaction Hamiltonian is linearized as $\hat{\mathcal{H}}_{\sf int} =  g \,\hat{x}_m \hat{x}_c$, and as usual we consider cavity decay rate $\kappa$, while the mechanical oscillator is coupled to a phononic Markovian bath characterized by $n_{\sf th}$ thermal phonons, and decoherence rate $\gamma$ \cite{RMPOptomechanics}.  A constant force is exerted on the mechanical oscillator, described by the Hamiltonian $\mathcal{\hat{H}}_\lambda = \lambda \hat{x}_m$, where $\lambda$ is the parameter we want to estimate. The details of the master equation for the two-mode density operator and of the corresponding Gaussian description can be found in Appendix \ref{appendix3}.\\
In order to estimate the force parameter $\lambda$, the {\em environment} of the cavity field, {\em i.e.} the cavity output field, is measured continuously as in \cite{KalmanOpto} (it is possible to include also the continuous measurement on the oscillator environment, which can be performed experimentally in certain opto-mechanical setups, for example by monitoring the light scattered from a levitating nanosphere \cite{Levante,Lia2016}).\\
Here we consider continuous homodyne detection with finite efficiency $\eta$, whose FI can be easily evaluated via the formulas presented in the previous example. It is important to notice how in this case, only the vector ${\bf u}$ depends on the parameter. As a consequence, since $\partial_\lambda A = 0$, also the matrix $\partial_\lambda \sigmaCM_t = 0$ at any time, and the evolution of the vector $\partial_\lambda {\bf R}_t$ is completely deterministic, reading
\begin{align}
\frac{d(\partial_\lambda {\bf R}_t )}{dt}= [A + (\sig_t B + N) B^{\sf T} ] (\partial_\lambda {\bf R}_t) + \partial_\lambda {\bf u}  \,.
\end{align}
It is then not necessary then to average the infinitesimal FI in Eq. (\ref{eq:FisherTCM}), and its value can be easily obtained numerically without needing to average over thousands of trajectories. We find that, as one could expect,  the optimal measurement is obtained for $\phi=\pi/2$, i.e. for homodyne monitoring of quadrature $\hat{p}_c$; moreover in Fig. \ref{f:forceest} we report the behaviour of the Fisher information as a function of time and for different values of the loss parameter $\kappa$ and we observe that the Fisher information is monotonically increasing with $\kappa$ for all the values of $\omega_m t$ we have investigated. Also this results is somehow expected as $\kappa$ represents in this picture the strength of the measurement performed via the environmental modes.\\
This example clearly shows the potential of our method: in fact, in order to evaluate the FI of this estimation problem with the method described in \cite{GammelmarkCRB}, it would have been necessary to integrate numerically a stochastic master equation, over around thousands trajectories, for two-mode operators, and thus corresponding to approximated matrices of dimension $(d_m d_c) \times (d_m d_c)$  ($d_m$ and $d_c$ being the  truncated dimensions of the Fock space for respectively the mechanical oscillator and the cavity field). \\

We want to remark that analytical solutions can also be obtained in a similar experimentally relevant scenarios. We report here the example of single-mode force estimation with time-continuous monitoring described by the following stochastic master equation, that for example describes continuous monitoring of a levitated nanosphere undergoing momentum diffusion in situations where the interaction with the cavity mode can be neglected \cite{Pflanzer2012,Calypso},
\begin{align}
d\varrho = - i \lambda [ \hat{p}, \varrho ] \,dt + \kappa \mathcal{D}[\hat{x}]\varrho \,dt + \sqrt{\eta \kappa} \mathcal{H}[\hat{x}]\varrho \, dw \,.
\end{align}
The covariance matrix of the conditional state evolves deterministically as
\begin{align}
\sig_t 
 &= \left(
\begin{array}{c c}
 \frac{1}{1+2 \eta \kappa t} &  0 \\
0 & 1 + 2 \kappa t 
\end{array}
\right)\, .
\end{align}
In order to evaluate the Fisher information corresponding to the estimation of the parameter $\lambda$, one only needs the vector $\partial_\lambda \langle \hat{x}\rangle_t$, whose evolution is described by the equation
\begin{align}
\frac{d (\partial_\lambda \langle \hat{x}\rangle_t)}{dt} =  -\frac{2 \eta \kappa}{1+2\eta\kappa t}  (\partial_\lambda \langle \hat{x}\rangle_t) - 1 \,,
\end{align}
and that can be analytically solved as
\begin{align}
\partial_\lambda \langle \hat{x}\rangle_t = -\frac{1+\eta\kappa t}{1+2 \eta\kappa t} t \,.
\end{align}
The corresponding infinitesimal and total Fisher information can be evaluated straightforwardly and one obtains
\begin{align}
dF_t (\lambda)&= \frac{4 t^2 \eta \kappa (1 + t \eta \kappa)^2}{(1+2 t \eta \kappa)^2}\, , \nonumber \\
F_t (\lambda)&= \frac{2 t^3 \eta \kappa (2 + t \eta \kappa)}{ 3 ( 1+ 2 t \eta \kappa)} \,.
\end{align}
In this case one can analytically check that $F_t(\lambda)$ is monotonically increasing with $\kappa$, and one also observes that, for large monitoring times, the Fisher information presents a remarkable $t^3$-scaling.

\section{Discussion and outlooks}\label{s:conclusions}
We have presented a reliable method for the calculation of the FI for dynamical parameter estimation in continuously measured linear Gaussian quantum systems. As shown in the two examples here described, our method greatly simplifies, in terms of computation complexity, the calculation of bounds on the estimation of parameters for such a rich and physically relevant family of quantum systems, compared to both the method presented in Ref. \cite{GammelmarkCRB} and the ones derived for general classical Gaussian systems \cite{Cavanaugh96,Klein2000A,Klein2000B,Ober2002}. It can also provide analytical results, allowing one to investigate in more detail the role played by the different physical parameters and to compare easily the efficiency of different measurement strategies. Furthermore, as Eqs. (\ref{eq:Kalman}) are formally equivalent to the classical continuous-time Kalman filter, our method can be generalized to the classical case (a more detailed discussion can be found in Appendix \ref{appendix4}).\\
Our results will find applications in assessing the performances of quantum sensors in several physical systems. It is worth mentioning the estimation of a magnetic field, in cases where an atomic spin ensemble can be approximated by a bosonic field via the Holstein-Primakovv approximation \cite{GeremiaMagnetometry2003}, and in several other quantum optomechanics setups and estimation problems, in particular with the aim of testing fundamental theories as corrections to Newtonian gravity \cite{Geraci} or to the Schroedinger equation \cite{RMPBassi,Calypso}. Moreover this approach can be generalized to the estimation of stochastic parameters, {\em e.g.} stochastic forces on mechanical oscillators, and of parameters characterizing the interaction with the environment and its temperature.\\

{\em 
Notice that this arXiv version of the manuscript includes the Erratum of the version published in Phys. Rev. A. 
}

\noindent
\section*{Acknowledgments} 
The author thanks S. Conforti, M. Paris and A. Serafini for discussions and constant support,  A. Mari for useful discussions regarding the additive property of the Fisher information and F. Albarelli for several discussions that contributed to find the error in the previous version of the manuscript.
The author acknowledges support from Marie Sk\l odowska-Curie Action H2020-MSCA-IF-2015 (project ConAQuMe, grant nr. 701154).
\appendix
\section{\label{appendix1}
Gaussian (general-dyne) measurements}
We here briefly present the parametrization of Gaussian measurements that are discussed and used in the article.\\
A Gaussian measurement is univocally described by a matrix $\sigmaCM_m$, s.t. $\sigmaCM_m + i \Omega \geq 0$ and the corresponding measurement outcomes are described by a vector ${\bf x}_m$. We start by focusing on single-mode projective Gaussian measurements, which thus correspond in the Hilbert space to projection onto a single-mode state $|\psi_G\rangle$. As the most general single-mode Gaussian state is a displaced squeezed vacuum state, the corresponding general matrix $\sigmaCM_m$ can be written as 
\begin{align}
\sigmaCM_m (s,\phi) &= R(\phi)
\left(
\begin{array}{ c c}
s & 0 \\
0 & 1/s 
\end{array}
\right) R(\phi)^{\sf T} \:,   \label{eq:projmeas}
\end{align}
with
\begin{align}
R(\phi) = 
\left(
\begin{array}{ c c}
\cos\phi & \sin\phi \\
-\sin\phi & \cos\phi 
\end{array}
\right) \:. 
\end{align}
In the case of homodyne measurement of the quadrature operator 
$\hat{r}_\phi = \cos\phi \, \hat{x} + \sin\phi\, \hat{p}$, one has to evalulate the limit
$\sigmaCM_m^{\sf (hom)} = \lim_{s\rightarrow 0} \sigmaCM_m (s,\phi)$, while for heterodyne detection, {\em i.e.} projection onto coherent states
one has $\sigmaCM_m^{\sf (het)} = \sigmaCM_m(1,\phi)$. \\
In order to take into account of inefficient detection, the measurement matrix $\sigmaCM_m^{(\eta)}$
is calculated via the action of
the dual noisy map on the {\em projective measurement} covariance matrix of Eq. (\ref{eq:projmeas}) as  \cite{diffusone}
\begin{align}
\sigmaCM_m^{\sf (ineff)} = X^* \sigmaCM_m X^{* {\sf T}} + Y^* \:, \label{eq:inefficient}
\end{align}
where
\begin{align}
X^* &= \mathbbm{1}_2 /\sqrt{\eta} \:, \nonumber \\
Y^* &= \frac{1-\eta}{\eta} \mathbbm{1}_2 \:, \nonumber
\end{align}
and $\eta$ quantifies the detection efficiency. \\
Notice that if one wants to consider a two-mode (or even multi-mode) local measurement, the overall measurement matrix $\sigmaCM_m$ is obtained by taking the direct sum of the single-mode measurements, {\em e.g.} $\sigmaCM_m = \sigmaCM_{m, (a)} \oplus \sigmaCM_{m, (b)}$. If one of the modes is not monitored, then one has to use the {\em inefficient} measurement matrix $\sigmaCM_m^{\sf (ineff)}$ as  in Eq. (\ref{eq:inefficient}) and take the limit for the corresponding efficiency parameter $\eta \rightarrow 0$.
\section{\label{appendix2} Diffusive and conditional evolution under time-continuous measurements}
In this section we will briefly describe the formalism and the calculations developed in Ref. \cite{diffusone} that lead to the equations describing the evolution of Gaussian states under time-continuous general-dyne measurements on the environment. \\
To keep the presentation easier, we consider the case where no Hamiltonian for the system is present, {\em i.e.} $\mathcal{H}_s=0$ and we focus on the interaction between the system and the bath degrees of freedom. This interaction is described by the Hamiltonian
\begin{align}
\mathcal{\hat{H}}_C = \hat{\bf r}^{\sf T} C \hat{\bf r}_{b}(t) = \frac12 \hat{\bf r}^{\sf T}_{sb} H_{C} \hat{\bf r}_{sb} =  
\frac12\hat{\bf r}^{\sf T}_{sb} \left(\begin{array}{cc}
0 & C \\
C^{\sf T} & 0
\end{array}\right) \hat{\bf r}_{sb} \, , \label{eq:Hc}
\end{align}
where $\hat{\bf r}_{b}(t)$ are the bath operators, defined by the white-noise condition
\begin{align}
\langle \{ \hat{\bf r}_b (t), \hat{\bf r}_b^{\sf T} (t') \} \rangle = \sigmaCM_b \delta(t-t') \:, \label{eq:SMwn}
\end{align}
and $\hat{\bf r}^{\sf T}_{sb} = (\hat{\bf r}^{\sf T},\hat{\bf r}_{b}(t)^{\sf T})$. From Eq. (\ref{eq:SMwn}) one notices 
that the operators $\hat{\bf r}_b(t)$
have the dimensions of the square root of a frequency. 
One can then define the so called {\em quantum Wiener increment} ${\rm d}\hat{\bf r}_b(t) $ as \cite{gardiner,WisemanMilburn}:
\begin{align}
{\rm d}\hat{\bf r}_b(t) = \hat{\bf r}_b(t) \, dt = \hat{\bf r}'_b(t) \,dW
\end{align}
and impose that $\hat{\bf r}_b'(t)$ is a vector of 
dimensionless field operators (that can be associated with {\em detector clicks} in the laboratory, and formally with POVM operators in the Hilbert space), satisfying the canonical commutation relations,
\begin{align}
[\hat{\bf r}'_{b}(t) , \hat{\bf r}'_b(t)] = i \Omega \:. \label{eq:rprime}
\end{align}
By observing that
\begin{align}
[{ d}\hat{\bf r}_b(t),{\rm d}\hat{\bf r}_b(t)] &=[\hat{\bf r}'_{b}(t)\, dW , \hat{\bf r}'_{b}(t) \,dW] = i \Omega \, dW^2 \:, \\
&=[\hat{\bf r}_{b}(t)\,dt , \hat{\bf r}_{b}(t) \,dt] = i \Omega\, dt \:,
\end{align}
one obtains the relationship $dW^2 = dt$. One can then interpret $dW$ as a stochastic Wiener increment, which is indeed responsible for the diffusive behaviour of the dynamics (in the Heisenberg picture the system and bath operators show in fact a random-walk like evolution). It is important to remark that the properties of $dW$ are a consequence of the white-noise condition describing the input operators $\hat{\bf r}_b(t)$ (other types of correlations would lead to a different stochastic behaviour). \\
We can now apply the Gaussian formalism to the vector of (well-defined) canonical operators $\hat{\bf r}'^{\sf T}_{sb} = (\hat{\bf r}^{\sf T},\hat{\bf r}'_{b}(t)^{\sf T})$. Under the coupling described in Eq. (\ref{eq:Hc}),  the dynamics over an interval $dt$ 
is generated by the operator $\hat{\bf r}^{\sf T}_{sb} H_{C} \hat{\bf r}_{sb}\,dt  = 
\hat{\bf r}^{\prime \sf T}_{sb} H_{C} \hat{\bf r}'_{sb}{dW}$. By expanding the corresponding symplectic transformation as
\begin{align}
{\rm e}^{\Omega H_{C} dW}  &\approx \left( \mathbbm{1} + \Omega{H}_{C}\,{dW} + \frac{(\Omega{H_{C}})^{2}}{2}\,{d}t \right) \,,
\end{align}
one calculates the evolution of the system-bath covariance matrix as
\begin{widetext}
\begin{align}
{\rm e}^{\Omega{H}_{C} dW} \left(\sig\oplus\sig_b\right) {\rm e}^{(\Omega{H_{C}})^{\sf T} dW}
 \approx&
\left(\sig\oplus\sig_b\right) + \left(A \sig+\sig A^{\sf T} + D\right)\oplus \tilde{\sig}_{b} \,dt
+ \sig_{sb}\, dW  \; , \label{eq:SMcov} 
\end{align}
\end{widetext}
where $A=(\Omega C \Omega C^{\sf T})/2$ and $D=\Omega{C} \sig_b {C}^{\sf T}\Omega^{\sf T}$ are typically addressed as drift and diffusion matrices, while the other matrices are
\begin{align}
\sig_{sb} &= \left(\begin{array}{cc} 
0 & \Omega{C} \sig_{b} + \sig {C\Omega^{\sf T}} \\
\sig_{b}{C}^{\sf T}\Omega^{\sf T} + \Omega{C^{\sf T}} \sig & 0
\end{array}\right) \:,\\
\tilde{\sig}_{b} &=  \frac{\Omega C^{\sf T} \Omega C \sig_b+\sig_b C^{\sf T}\Omega C \Omega}{2} + \Omega^{\sf T} C^{\sf T} \sig C \Omega \:.
\end{align}
The same procedure can be applied in order to obtain the evolution of the first moments, s.t.
\begin{align}
{\rm e}^{\Omega H_{C} dW} 
\left(\begin{array}{c}
{\bf R}  \\
{\bf R}_b
\end{array}\right)
&\approx \left( \mathbbm{1} + \Omega{H}_{C}{dW} + \frac{(\Omega{H_{C}})^{2}}{2}\,{\rm d}t \right)
\left(\begin{array}{c}
{\bf R} \\
{\bf R}_b
\end{array}\right)  \\
&\approx
\left(\begin{array}{c}
{\bf R}  + A {\bf R} \,dt \\
\Omega C^{\sf T} {\bf R} \, dW
\end{array}\right)\, \label{eq:SMfm} .
\end{align}
In the last equation we have assumed ${\bf R}_b=0$, {\em i.e.} that the bath operators have first moments equal to zero.\\
The probability density for the outcomes ${\bf x}_m$ of a general-dyne measurement on the bath degrees of freedom, described by the covariance matrix $\sig_m$, is a multivariate Gaussian distribution
\be
p({\bf x}_m) = \frac{e^{- \frac{1}{2}({\bf x}_m - \bar{\bf x}_m)^{\sf T} \boldsymbol{\Sigma}^{-1} ({\bf x}_m - \bar{\bf x}_m)}}{(2 \pi)^n \sqrt{{\rm Det}[\boldsymbol{\Sigma}]}}
\ee
with first moments vector $\bar{{\bf x}}_m =\Omega C^{\sf T} {\bf R} \, dW$, and with covariance matrix $\boldsymbol{\Sigma} = (\sig_b + \sig_m)/2$. One can then define the vector of random variables
\begin{align}
d{\bf w} = \left(\frac{\sig_b + \sig_m}{2}\right)^{-1/2} ( {\bf x}_m - \bar{\bf x}_m )\, dW \,
\end{align}
and, by exploiting the statistical properties of the measurement outcomes ${\bf x}_m$, one observe that  $d{\bf w}$ is in fact a vector of uncorrelated real Wiener increments, {\em i.e.} it is distributed according to a Gaussian with first moments $\mathbbm{E}[d{\bf w}]=0$ and covariance matrix $\mathbbm{E}[\{ d{\bf w} , d{\bf w}^{\sf T} \}/2] = \mathbbm{1} dt$. 
It is often customary to express the measurement outcomes as a real current with uncorrelated noise \cite{WisemanDoherty}:
\begin{align}
d{\bf y} &:= (\sigmaCM_b + \sigmaCM_m)^{-1/2} \:{\bf x}_m \,dW \nonumber \\ 
 &= (\sigmaCM_b + \sigmaCM_m)^{-1/2} \bar{\bf x}_m \,dW +\frac{\bf dw}{\sqrt{2}} \nonumber \\
 &= (\sigmaCM_b + \sigmaCM_m)^{-1/2} \Omega C^{\sf T} {\bf R}_t \,dt +\frac{\bf dw}{\sqrt{2}} \,.
\end{align}

Finally, starting from Eqs. (\ref{eq:SMcov}) and (\ref{eq:SMfm}), and by applying the formulas for the evolution of Gaussian states under conditional measurements on one of the two subsystems, one obtains the equations \cite{diffusone}
\begin{align}
d{\bf R}_t &= A {\bf R}_t \, dt + {\bf u} \, dt + \left(\frac{\sigmaCM_t B + N}{\sqrt{2}}\right) {\bf dw}\:, \nonumber \\
\frac{d\sigmaCM_t}{dt} &= A \sigmaCM_t + \sigmaCM_t A^{\sf T} + D -  (\sigmaCM_t B + N ) (\sigmaCM_t B + N )^{\sf T} \:. \label{eq:stocast} 
\end{align} 

\section{\label{appendix3} Estimation of a constant force on a mechanical oscillator}
We consider a standard cavity optomechanical setup where a mechanical oscillator, described by position and momentum operators $\hat{x}_m$ and $\hat{p}_m$ and oscillating at frequency $\omega_m$, is coupled to a cavity mode, described by amplitude and phase operators $\hat{x}_c$ and $\hat{p}_c$ and with resonance frequency $\omega_c$. Assuming that the cavity is strongly driven with a laser at frequency $\omega_l$, one can consider the following linearized Hamiltonian in a frame rotating with $\omega_l$: 
\begin{align}
\mathcal{\hat{H}}_{\sf om} = \omega_m (\hat{x}_m^2 + \hat{p}_m^2)/2  - \Delta ( \hat{x}_c^2 + \hat{p}_c^2)/2 + g \,\hat{x}_m \hat{x}_c \:,
\end{align}
where $\Delta = \omega_l -\omega_c$ is the detuning of the driving laser with respect to the cavity, and $g$ is the effective optomechanical coupling strength. If a constant force is exerted on the mechanical oscillator, then one has to add the linear Hamiltonian $\mathcal{\hat{H}}_\lambda = \lambda \hat{x}_m$. We also assume that the cavity has a decay rate $\kappa$, while the mechanical oscillator is coupled to a phononic Markovian bath characterized by $n_{\sf th}$ thermal phonons, and the corresponding width of the mechanical resonance is equal to $\gamma$. The master equation for the two-mode density operator reads
\begin{align}
\frac{d\varrho}{dt} &= - i [ \mathcal{\hat{H}}_{\sf om}+\mathcal{\hat{H}}_\lambda, \varrho] + \kappa \mathcal{D}[\hat{a}] \varrho  + 
\gamma (n_{\sf th} + 1) \mathcal{D}[\hat{b}] \varrho  \nonumber \\
& \:\:\: +  \gamma n_{\sf th} \mathcal{D}[\hat{b}^\dag] \varrho \:,
\end{align}
where $\hat{a} = (\hat{x}_c + \hat{p}_c)/\sqrt{2}$ and $\hat{b} = (\hat{x}_m + \hat{p}_m)/\sqrt{2}$.\\

In the Gaussian picture the interaction with the environment and its correlations are described by the matrices
\begin{align}
C &= \left(
\begin{array}{c c c c}
0 & -\sqrt{\kappa} & 0 & 0 \\
\sqrt{\kappa} & 0 & 0 & 0 \\
0 & 0 & 0 & \sqrt{\gamma} \\
0 & 0 & -\sqrt{\gamma}  & 0 
\end{array}
\right) \nonumber \\
\sigmaCM_b &= \left(
\begin{array}{c c c c}
1 & 0 & 0 & 0 \\
0 & 1 & 0 & 0 \\
0 & 0 &  1 + 2 n_{\sf th} & 0 \\
0 & 0 & 0 & 1 + 2 n_{\sf th}
\end{array}
\right). \nonumber
\end{align}
By also assuming that a continuous homodyne measurement with efficiency $\eta$ is performed on the environment (in this case on the cavity output field), the dynamics is equivalently described by the equations for the first moment vector and for the covariance matrix (\ref{eq:stocast}). 
The corresponding matrices and vectors read
\begin{align}
A &= \left(
\begin{array}{c c c c}
-\kappa/2 & -\Delta & 0 & 0 \\
\Delta & -\kappa/2 & -g & 0 \\
0 & 0 & -\gamma/2 & \omega_m \\
-g & 0 & -\omega_m & -\gamma/2 
\end{array}
\right) \nonumber \\
D &= \left(
\begin{array}{c c c c}
\kappa & 0 & 0 & 0 \\
0 & \kappa & 0 & 0 \\
0 & 0 & \gamma ( 1 + 2 n_{\sf th}) & 0 \\
0 & 0 & 0 & \gamma(1 + 2 n_{\sf th})
\end{array}
\right) \nonumber \\
B &= - N = \left(
\begin{array}{c c c c}
\sqrt{\eta \kappa} \cos^2{\phi} & \sqrt{\eta \kappa} \sin\phi \cos{\phi}  & 0 & 0 \\
-\sqrt{\eta \kappa} \sin\phi \cos{\phi}  & \sqrt{\eta \kappa} \sin^2{\phi}  & 0 & 0 \\
0 & 0 & 0 & 0 \\
0 & 0 & 0 & 0
\end{array}
\right) \nonumber \\
{\bf u}^{\sf T} &= \left( 0 \: 0\: 0 \: -\lambda \right). \nonumber 
\end{align}
\noindent
As only the vector ${\bf u}$ depends on the parameter $\lambda$, with $\partial_\lambda {\bf u}^{\sf T}=(0 \, 0\, 0, -1 )$, in order to calculate the infinitesimal FI, one has only to solve the following differential equation for $(\partial_\lambda {\bf R}_t)$:
\begin{align}
\frac{d(\partial_\lambda {\bf R}_t )}{dt}= [A + (\sig_t B + N) B^{\sf T} ] (\partial_\lambda {\bf R}_t) + \partial_\lambda {\bf u}  \,.
\end{align}
\section{\label{appendix4}
Generalization to the classical continuous-time Kalman filter}
The equations (\ref{eq:Kalman}) describing the Gaussian quantum dynamics are in fact formally equivalent to the classical continuous-time Kalman filter. As a consequence we can generalize the method to the classical case, obtaining similar equations as the ones presented in Refs.\cite{Cavanaugh96,Klein2000A,Klein2000B,Ober2002}.\\
The goal of the Kalman filter is to obtain an estimate
 ${\bf \hat{x}}_t$ of a certain process ${\bf x}_t$, via a continuous measurement output ${\bf y}_t$ with uncorrelated noise. The corresponding equations read
\begin{align}
d{\bf \hat{x}}_t &= A {\bf \hat{x}}_t \, dt + {\bf u} \, dt + \left({\boldsymbol \Sigma}_t B + N \right) {\bf dw}\:, \nonumber \\
\frac{d{\boldsymbol \Sigma}_t}{dt} &= A {\boldsymbol \Sigma}_t + {\boldsymbol \Sigma}_t A^{\sf T} + D -  ({\boldsymbol \Sigma}_t B + N ) ({\boldsymbol \Sigma}_t B + N )^{\sf T} \:, \nonumber \\
d{\bf y}_t &= B^{\sf T} \hat{\bf x}_t \,dt + {\bf dw} \:, \label{eq:classicKalman} 
\end{align}
where ${\boldsymbol \Sigma}_t$ represents the mean squared error matrix of the estimate.
As the increment of the time continuous output $d{\bf y}_t$ is a Gaussian random variable with mean value vector $\langle d{\bf y}_t \rangle = B^{\sf T} \hat{\bf x}_t \, dt$ and covariance matrix ${\bf \Gamma}_t = \mathbbm{1} dt$, the corresponding infinitesimal FI (for a specific {\em trajectory}) can be calculated by using Eq. (\ref{eq:FisherGauss}).  In particular one obtains 
\begin{align}
dF_t^{\sf (traj)} (\theta) = \left[\partial_\theta (B^{\sf T}\hat{\bf x}_t) \right]^{\sf T} \left[\partial_\theta (B^{\sf T}\hat{\bf x}_t )\right] dt \:,
\end{align}
where the evolution of the vector $\partial_\theta (B^{\sf T}\hat{\bf x}_t )$ can be derived as in Eqs. (\ref{eq:derivatives}). In order to evaluate the FI for the classical continuous-time Kalman filter,
corresponding to the whole stream of outcomes up to time $t$, one can then average $dF_t^{\sf (traj)}(\theta)$ over the Wiener process and perform the integral over time as in Eqs. (\ref{eq:average}) and (\ref{eq:integral}).
%
%
%

\begin{thebibliography}{99}
\bibitem{noteExtension} We remark that all the results we will present in the following can be generalized to the case where the parameter $\theta$ characterizes the dissipative dynamics, either via the interaction between the system and the environment, or via the correlations of the environment itself.
\end{thebibliography}
\end{document}